\documentclass{article}

\usepackage{arxiv}

\usepackage[numbers]{natbib}
\usepackage{hyperref}       
\usepackage[utf8]{inputenc} 
\usepackage[T1]{fontenc}    
\usepackage{url}            
\usepackage{booktabs}       
\usepackage{amsfonts}       
\usepackage{nicefrac}       
\usepackage{microtype}      
\usepackage{lipsum}         
\usepackage{graphicx}
\usepackage{doi}
\usepackage{amsmath}
\usepackage{algorithm}
\usepackage{algpseudocode}
\usepackage{amsfonts}
\usepackage{hyperref}
\usepackage[export]{adjustbox}
\usepackage{cleveref}       

\title{Derivatives Sensitivities Computation under Heston Model on GPU}

\author{Pierre-Antoine Arsaguet\\
	Department of Computing\\
	Imperial College London\\
	South Kensington Campus\\
    Exhibition Road\\
    London SW7 2AZ\\
	\texttt{pantoinea@gmail.com} \\
	\And
	\href{https://orcid.org/0000-0001-6846-6649}{\includegraphics[scale=0.06]{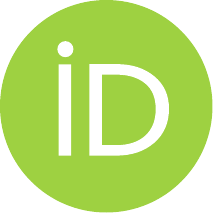}\hspace{1mm}Paul Alexander Bilokon} \\
	Department of Computing\\
    Department of Mathematics\\
	Imperial College London\\
	South Kensington Campus\\
    Exhibition Road\\
    London SW7 2AZ\\
	\texttt{paul.bilokon@imperial.ac.uk} \\
}

\date{27 August, 2023}

\begin{document}
\maketitle

\begin{abstract}
    This report investigates the computation of option Greeks for European and Asian options under the Heston stochastic volatility model on GPU. We first implemented the exact simulation method proposed by Broadie and Kaya and used it as a baseline for precision and speed. We then proposed a novel method for computing Greeks using the Milstein discretisation method on GPU. Our results show that the proposed method provides a speed-up up to 200x compared to the exact simulation implementation and that it can be used for both European and Asian options. However, the accuracy of the GPU method for estimating Rho is inferior to the CPU method. Overall, our study demonstrates the potential of GPU for computing derivatives sensitivies with numerical methods.
\end{abstract}

\section{Introduction}

The efficient computation of financial derivatives prices and sensitivities, also known as Greeks, is crucial for financial institutions to manage risks and make informed investment decisions. The Heston model \cite{RePEc:oup:rfinst:v:6:y:1993:i:2:p:327-43} is a widely used stochastic volatility model that accurately captures the dynamics of asset prices in financial markets. However, the computation of option prices and Greeks under this model can be computationally expensive, especially when dealing with complex options and large portfolios.

In recent years, graphics processing units (GPUs) have emerged as a powerful tool for accelerating numerical computations in various fields, including finance. In this project, we investigate the potential of using GPUs to compute option prices and Greeks under the Heston model. Since financial computation mainly relies on Monte Carlo methods, GPUs are particularly good at parallelising such simulations and thus accelerating them. Specifically, we focus on implementing the Exact Simulation method and the Milstein discretisation method on GPUs.

We compare the performance and accuracy of the GPU implementations with the corresponding CPU implementations. We also evaluate the impact of various parameters, such as the number of simulations and discretisation steps, on the performance and accuracy of the GPU implementations. Furthermore, we investigate the potential of using quasi-Monte Carlo methods to further improve the accuracy of the GPU implementations.

Finally, we compute option Greeks using the GPU implementations and compare the results with those obtained from the CPU implementations and with analytical values. Our results demonstrate the potential of using GPUs for efficient and accurate computation of option prices and Greeks under the Heston model, which can have important applications in risk management and investment decision-making.

This work follows the pattern of~\cite{WILM:WILM11171}, in which Quasi-Monte Carlo was applied to compute the Greeks for vanilla options under the Heston model.

\section{Background}

Derivatives are financial instruments whose value is derived from an underlying asset, such as a stock, bond, commodity, or currency. The underlying asset can be almost anything, but the most commonly traded derivatives are based on stocks and stock indices. Derivatives are used by investors to manage risk, speculate on future price movements, or gain exposure to different markets and assets. The most basic type of derivative is a forward contract, which is an agreement to buy or sell an asset at a specified price on a future date. However, the most commonly traded derivatives are options.

An option is a type of derivative that gives the holder the right, but not the obligation, to buy or sell an underlying asset at a specified price on or before a specified date. There are two main types of options: call options and put options. A call option gives the holder the right to buy an underlying asset at a specified price, while a put option gives the holder the right to sell an underlying asset at a specified price. The price at which the underlying asset can be bought or sold is known as the strike price, and the specified date is known as the expiration date. The price of an option is determined by a variety of factors, including the price of the underlying asset, the volatility of the asset, and the time to expiration.

European options are a type of option where the holder can exercise the option only on the expiration date. The payoff of a European call option with strike price $K$ and underlying asset price $S_T$ at expiration is given by $C_{euro} = \max(S_T-K,0)$. Similarly, the payoff of a European put option is given by $P_{euro} = \max(K-S_T,0)$.

Asian options are a type of option where the payoff is determined by the average price of the underlying asset over a certain period of time, rather than the price at expiration. A commonly used type of Asian option is the arithmetic average Asian option, where the payoff is based on the average price of the underlying asset over a fixed time period. The payoff of an arithmetic average Asian call option with strike price $K$ is defined by the average price over $N$ times $(t_i)_{1 \leq i \leq N}$ $\bar{S}_T = \frac{1}{N} \sum^N_{i=1} S_{t_i}$ is given by $C_{asian} = \max(\bar{S}_T-K,0)$. Similarly, the payoff of an arithmetic average Asian put option is given by $P_{asian} = \max(K-\bar{S}_T,0)$.

Note that the average price $\bar{S}_T$ is typically calculated using either the arithmetic mean or the geometric mean of the underlying asset prices over the specified time period.

Option pricing under the risk-neutral measure is a popular framework used in finance to price options. In this framework, the price of an option is derived by discounting the expected payoff of the option by a risk-free interest rate. The risk-neutral measure assumes that the expected return of the underlying asset is equal to the risk-free interest rate. Therefore, the expected payoff of the option can be computed using a risk-neutral probability measure, which is a probability measure under which the discounted price of the underlying asset is a martingale. The risk-neutral probability measure is not the same as the real-world probability measure, but it allows us to price options as if there were no risk in the underlying asset. The option price under the risk-neutral measure is then the discounted expected payoff of the option, where the discount factor is the risk-free interest rate.

Suppose that we have an underlying asset with price process $S_t$, and we consider an option with payoff $h(S_T)$ at time $T$. The price of the option at time $t$ is given by the risk-neutral expectation of the discounted payoff, that is, $V(t,S_t) = e^{-r(T-t)}\mathbb{E}_{\mathbf{Q}}\left[ h(S_T)\ |\ S_t \right],$ where $r$ is the risk-free interest rate, and $\mathbf{Q}$ is the risk-neutral probability measure. Under the risk-neutral measure, the discounted asset price $e^{-rt}S_t$ is a martingale. Therefore, we can compute the option price as the discounted expected payoff, where the expectation is taken under the risk-neutral probability measure $\mathbf{Q}$. Then, we need a model to estimate the asset price at time T. This could be done by models such as Black-Scholes model, or Heston model, which is used for this work. 

Derivatives sensitivities, or ``Greeks'' are a set of risk measures that are used to assess the sensitivity of an option's price to changes in various parameters, such as the underlying asset price, volatility, and time to expiration. These risk measures are important tools for option traders and portfolio managers, as they provide insight into the behavior of an option under different market conditions. The most commonly used Greeks are Delta, Gamma, Vega, Theta, and Rho. In this report, we will focus on Delta, Gamma and Vega. Delta measures the sensitivity of an option's price to changes in the underlying asset price, while Gamma measures the rate of change of Delta. Vega measures the sensitivity of an option's price to changes in volatility. Given an option with a price $V$, based on an asset $S$ which underlying volatility is $\sigma$, Greeks formulas are given by $\text{Delta} = \frac{\partial V}{\partial S}$, $\text{Gamma} = \frac{\partial^2 V}{\partial S^2}$, $\text{Rho} = \frac{\partial V}{\partial \sigma}$.

\subsubsection{Pathwise method for greeks estimation}
Computation of Greeks is time consuming, thus approximation algorithms have been designed for an efficient Greeks calculation. A popular method is the Pathwise method, developed by Glasserman in 1996 \cite{Broadie1996EstimatingSP}, later explained and applied to Stochastic Volatility by Broadie and Kaya~\cite{inproceedings}. While being a unbiased estimator of Greeks, the Pathwise estimator is also computationally efficient. Mathematical details of the Pathwise method would be beyond this work, however, we can explain the intuition behind it. Take a European option, whose payoff value is
\begin{equation}
    V = e^{-rT} \max(K-S(T),0) 
\end{equation}
We can illustrate the Pathwise method for $\Delta$. Thanks to the chain rule, we have:
\begin{equation}
    \frac{\partial V}{\partial S(0)} = \frac{\partial V}{\partial S(T)} \frac{\partial S(T)}{\partial S(0)}
\end{equation}
Because we have $\frac{\partial S(T)}{\partial S(0)} = \frac{S(T)}{S(0)}$ and $\frac{\partial V}{\partial S(T)} = e^{-rT} \mathbf{1} \{S(T)>K\} $, we can state:
\begin{equation}
    \label{eq:pw_est}
    \frac{\partial V}{\partial S(0)} = e^{-rT} \frac{S(T)}{S(0)} \mathbf{1} \{S(T)>K\} 
\end{equation}

Equation \ref{eq:pw_est} is an unbiased estimator of $\Delta$, and is easily computed once we have $S(T)$. One condition for such a Greek estimation is that the payoff function must be Lipschitz continuous. Broadie and Kaya derived Pathwise estimators for Stochastic Volatily models for European options: 
\begin{align}
    \label{eq:euro_pw}
    & \text{Delta} = e^{-rT} \frac{S(T)}{S(0)} \mathbf{1} \{S(T)>K\} \\
    & \text{Rho} = e^{-rT} K T \mathbf{1} \{S(T)>K\} 
\end{align}

While Arithmetic Asian option Greeks are given by:
\begin{align}
    \label{eq:asian_pw}
    & \text{Delta} = e^{-rT} \frac{\bar{S}}{S(0)} \mathbf{1} \{\bar{S}>K\} \\
    & \text{Rho} = e^{-rT} K T \mathbf{1} \{\bar{S}>K\} \left ( \frac{1}{N} \sum^{N}_{i=1} S_{t_i} t_i - T(\bar{S} - K)  \right) 
\end{align}

\subsection{Heston Model for option pricing}

The Heston model is a popular stochastic volatility model used in finance to price derivatives. It was introduced by Steven Heston in 1993 \cite{RePEc:oup:rfinst:v:6:y:1993:i:2:p:327-43}, and it is widely used to model asset prices that exhibit stochastic volatility. The model assumes that the volatility of the underlying asset follows a stochastic process that is driven by a Brownian motion. The model is expressed in terms of two stochastic differential equations (SDEs), one for the asset price and the other for the volatility process. For a given asset $S$ with a volatility $V$, the SDEs can be written in the following mathematical form:
\begin{align}
\label{eq:heston_eq}
\mathrm{d}S_t &= r S_t \mathrm{d}t + \sqrt{v_t} S_t \mathrm{d}W^{(1)}_{t}\\
\mathrm{d}V_t &= \kappa(\theta-V_t) \mathrm{d}t + \sigma \sqrt{V_t} \mathrm{d}W^{(2)}_{t}
\end{align}
where: 
\begin{itemize}
\item $r$ is the risk-free rate 
\item $\theta$ is the average level of variance process V 
\item $\kappa$ is the mean speed of reversion
\item $\sigma$ is a parameter for the volatility model
\item $\mathrm{d}W^{(1)}_{t}, \mathrm{d}W^{(1)}_{t}$ are two Brownian motions with a correlation coefficient $\rho$
\end{itemize}

Two main approaches exist to simulate an asset price $S_t$, either get a sample from the actual distribution of St through exact simulation introduced by Broadie and Kaya and derived methods, or using an Euler approximation in the Heston equations \ref{eq:heston_eq}.  

\subsection{Heston Model Exact simulation}

Given initial conditions $S_u$, $V_u$, the integrated form of Equations \ref{eq:heston_eq} is: 
\begin{align}
S_t &= S_u \exp \left( r(t-u) - \frac{1}{2} \int^{t}_{u} V_s ds + \int^{t}_{u} \sqrt{V_s} dW^{(1)}_s \right) \\
V_t &= V_u + \kappa \theta (t-u) - \kappa \int^t_u V_s ds + \sigma \int^t_u \sqrt{V_s} dW^{(2)}_s \label{eq:broadie_alg}
\end{align}

Broadie and Kaya proposed an algorithm that conditions $S_t$ on variance process $V_t$ in order to get a sample of an asset price \cite{ExactBroadie}.

\begin{algorithm}
\label{alg:exact}
\caption{Broadie and Kaya exact simulation algorithm}
1. Generate a sample from the distribution of $V_t$ given $V_u$ \\
2. Generate a sample from the distribution of  $\int^t_u V_s ds$ given $V_t$ and $V_u$ \\
3. Recover $\int^t_u \sqrt{Vs} dW^{(2)}_s$ from \ref{eq:broadie_alg} given $V_t$, $ V_u$ and $\int^t_u V_s ds$ \\
4. Generate a sample from the distribution of $S_t$ given $\int^t_u \sqrt{Vs} dW^{(1)}_s$ and $\int^t_u V_s ds$ 
\end{algorithm}

Each step of this algorithm is detailed in the Broadie and Kaya \cite{ExactBroadie}. Nevertheless, this process contains steps that can make the simulation inefficient, especially on GPU, which is the objective of this work. Thus, only these steps will be detailed. Many researches have tackled the lack of efficiency and proposed new methods to sample from exact distributions.

\subsection{Exact simulation bottlenecks}
\subsubsection{Sampling $V_t$ from $V_u$}
Given $V_u$, the $V_t$ expression is: 
\begin{align}
V_t &= \frac {\sigma^2 (1- e^{-\kappa(t-u)})}{4 \kappa} \chi^{'2}_d \left (\frac{4 \kappa e^{- \kappa (t-u)}}{\sigma^2 (1- e^{-\kappa(t-u)})} \right )
\end{align}

For computational efficiency, especially when parallelising, it is important to use optimized methods to generate random distributions. When $d$ is an integer, one can derive a Non Central Chi Squared (NCCS) distribution from normal random variables. We recall that given ($Z_1, Z_2, ..., Z_i, ..., Z_{d-1}, Z_d$), $k$ independent Normal normal variables, with unit variances means ($\mu_1, ..., \mu_i, ..., \mu_{d-1}, \mu_d$) and the random variable G: 

\begin{align}
X &= \sum^d_1 Z^2_i
\end{align}
follows a Non Central Chi Squared distribution, with a degree of freedom $d$, and a Non Central Parameter $\lambda$ :

\begin{align}
\lambda &= \sum^d_1 \mu^2_i
\end{align}

However, since the degree of freedom $d$ is not an integer, a Non Central Chi Squared variable can not been directly derived from Normal variables. Nevertheless, when $d>1$, one can write: 

\begin{align}
\label{eq:chi2}
\chi^{'2}_d(\lambda) &=\chi^{2}_{d-1} + (Z + \sqrt{\lambda})^2
\end{align}

Modern programming languages, such as C++ and CUDA, which are used for the computations in this project, have built-in optimized routines to sample from normal random variables but not from chi-squared distributions. To compute an NCCS on CUDA, equation \ref{eq:chi2} must be transformed. A non-central chi-squared distribution is a particular case of a Gamma distribution:

\begin{align}
\chi^{2}_d & \sim \Gamma (\frac{d}{2}, 2)
\end{align}

Thus, equation \ref{eq:chi2} becomes: 

\begin{align}
\chi^{'2}_d(\lambda) &=\Gamma (\frac{d-1}{2}, 2) + (Z + \sqrt{\lambda})^2
\end{align}

In~\cite{GammaGeneration}, Marsalia and Tsang introduce a recursive algorithm to generate gamma variables only from Uniform and Normal random variables. Finally, the sample of $V_t$ given $V_u$ is feasible on both CPU and GPU in a optimized manner.

\subsubsection{Sampling from the distribution of  $\int^t_u V_s ds$}

If denote $V(t,u)$ a random variable having the same distribution as $\int^t_u V_s ds$ given $V_t$, $V_u$, Feller \cite{feller-vol-2} proves that: 

\begin{align}
Pr \left ( V(t,u)) \leq  x \right ) &= \frac {2}{\pi} \int^{\infty}_0 \frac{\sin{sx}}{s} \text{Re} [\Phi(s)]ds
\end{align}

Here, $\Phi(a)$ is defined as:

\begin{equation} 
\begin{split}
\Phi(a) = & \frac{\gamma(a) e^{-0.5 (\gamma(a) - \kappa)(t-u)}(1 - e^{\kappa (t-u})}{\kappa (1 - e^{ -\gamma(a) (t-u)})} \\
 & \times \exp{ \left ( \frac{V_u + V_t}{\sigma^2}\left[ \frac{\kappa (1+e^{\kappa (t-u)})}{1-e^{-\kappa (t-u)}} - \frac{\gamma(a) (1 + e^{ -\gamma(a) (t-u)}}{(1 - e^{ -\gamma(a) (t-u)}} \right] \right )} \\
 & \times \frac{I_{0.5d-1} \left[\sqrt{V_t V_u} \frac{4 \gamma(a) e^{-0.5 \gamma(a)(t-u)}}{\sigma^2 (1-e^{-\gamma(a)(t-u)}} \right ]} {I_{0.5d-1} \left[\sqrt{V_t V_u} \frac{4 \kappa e^{-0.5 \kappa(t-u)}}{\sigma^2 (1-e^{-\kappa(t-u)}} \right ]}
\end{split}
\end{equation}

where $\gamma(a) = \sqrt{\kappa^2 -2 \sigma^2 i a}$ and $I_v(z)$ denotes a Modified Bessel First Kind function with a complex argument. While C++ and CUDA have builtin Bessel functions for real argument, we need to use a routine such as the one provided by Amos \cite{Amos} for complex arguments. This routine introduces some serious complexity to the whole simulation process, especially if we want to parallelise it.

Once we computed $\Phi(a)$, the integral $\frac {2}{\pi} \int^{\infty}_0 \frac{\sin{sx}}{s} \text{Re} [\Phi(s)]ds$ can be approximated by different methods, such as trapezoidal integration approximation. Since the objective is to sample $V(t,u)$, and we only know the distribution $Pr \left ( V(t,u)) \leq  x \right )$, we can use the inverse transform method. For a given a random variable X with cumulative distribution $F_X(x) = Pr(X\leq x)$, the random variable $F^{-1}_X(U)$, with $U \sim \text{Unif}[0,1]$ has the same law as $X$. Thus, it is possible to sample from $V(t,u)$ distribution, provided that we can compute $F^{-1}_X(u)$ for any $u \in [0,1]$. 
The computation of $F^{-1}_X(u), u \in [0,1]$ could be done by the use of bisection methods, where the objective is to find $x$ such as $F_X(x) - u = 0$ for a given $u \in [0,1]$.  
Since the first and second derivatives of $Pr \left ( V(t,u)) \leq  x \right )$ are easily found, we can use second order Newton algorithm, where at each step the next value computed is:

\begin{align}
    x_{n+1} & = x_n - \frac{f^{'}(x_n)}{f^{''}(x_n)} \left ( 1 - \sqrt{\frac{2 f(x_n) f^{''}(x_n)}{f^{'}(x_n)^{2}}} \right )
\end{align}

Thus, many iterations have to be done to have a precise approximation of $V(u,t)$, and at each iteration, we need to compute $Pr \left ( V(t,u)) \leq  x \right )$ and its two first derivatives, for which an approximation of integral has to be made. Additionally, computing the Bessel function through a routine and sampling from $\int^t_u V_s ds$ is computationally expensive, making the exact simulation ineffective. Many alternatives have been proposed since then \cite{NewSimu}, but the most straightforward is the Euler discretization of the Heston model.

\subsection{Approximate simulation through discretisation}
\subsubsection{Euler discretisation}

The use of numerical methods to simulate stochastic processes has become increasingly important in financial engineering. One popular technique for approximate simulation is Euler discretisation. Euler's method proposes an approximation of the integral \ref{eq:int_euler}, with the left-hand and right-hand rectangle methods approximating the integral as $f(t)dt$ and $f(t+dt)dt$, respectively.

\begin{align}
    \label{eq:int_euler}
    \int^{t+dt}_t f(x)dx \approx f(t)dt
\end{align}

In the context of simulating financial processes, $f(t)$ could represent the stock price $S_t$ or the variance $V_t$, thus we are forced to use left hand method because $V_{t+dt}$ and $S_{t+dt}$ are unknown.

Under Euler's approximation, given $S_t$ and $V_t$, the stock price $S_{t+dt}$ and variance $V_{t+dt}$ can be estimated using the following equations:
\begin{equation} 
\begin{split}
S_{t+dt} &= S_t \exp \left( rdt - \frac{1}{2} \int^{t+dt}_{t} V_s ds + \int^{t+dt}_{t} \sqrt{V_s} dW^{(1)}_s \right) \\
& \approx S_t \exp \left (rdt - \frac{1}{2}V_t dt + \sqrt{V_t} (W^{(1)}_{t+dt} - W^{(1)}_t) \right )
\end{split}
\end{equation}

For a Brownian motion, $W^{(1)}_t - W^{(1)}_u \sim \text{N}(0,dt)$. Thus, if $Z_1 \sim \text{N(0,1)}$, we can write:
\begin{align}
S_{t+dt} & \approx S_t \exp \left ( (r - \frac{1}{2}V_t)dt + \sqrt{V_t} \sqrt{dt} Z_1 \right )
\end{align}

In the same way: 
\begin{equation} 
\begin{split}
V_{t+dt} &= V_t + \kappa \theta dt - \kappa \int^{t+dt}_t V_s ds + \sigma \int^{t+dt}_t \sqrt{Vs} dW^{(2)}_s \\
& \approx V_t + \kappa (\theta - V_t) dt + \sigma \sqrt{V_t} \sqrt{dt} Z_2
\end{split}
\end{equation}

Where $Z_1, Z_2$ are correlated by $\rho$. To generate two correlated Normal random variable, one can use the algorithm \ref{alg:corr_var}. 

\begin{algorithm}
\label{alg:corr_var}
\caption{Generation of two correlated Normal random variables}
1. Generate $Z_a, Z_b$ two normally distributed independent variables. \\
2. $Z_1 = Z_a$ \\
3. $Z_2 = \rho Z_1 + \sqrt{1-\rho^2} Z_b$
\end{algorithm}

Finally, although Euler discretisation is less precise, it has the advantage of being computationally efficient and easily parallelisable. A large number of Euler simulations can be performed to obtain a reasonable Monte Carlo approximation of the asset price. However, it should be noted that Euler's method still has a high bias and therefore, improved discretisation methods have been proposed, such as Milstein discretisation.

\subsubsection{Milstein discretisation}
Milstein proposed in 1971 \cite{Milstein} a discretisation method for Stochastic Volatility. Frouah \cite{FrouahWebsite} detailed it for Heston model. It applies to assets $S$ where we can make the assumption that $\mu, \sigma$ only depends on $S_t$: 

\begin{align}
\mathrm{d}S_t &= \mu(S_t) \mathrm{d}t + \sigma(S_t)\mathrm{d}W_t
\end{align}

Then, according to Ito's Lemma, we can write: 
\begin{equation} 
\begin{split}
\mathrm{d}\mu(S_t) &= \frac{\mathrm{d}\mu (St)}{\mathrm{d}x}\mathrm{d}S_t + \frac{1}{2} \frac{\mathrm{d}\mu(S_t)}{\mathrm{d}x^2} \sigma(S_t)^2 \mathrm{d}t \\
& = \frac{\mathrm{d}\mu (S_t)}{\mathrm{d}x}(\mu(S_t)\mathrm{d}t + \sigma(S_t)\mathrm{d}W_t) +  \frac{1}{2} \frac{\mathrm{d}\mu(S_t)}{\mathrm{d}x^2} \sigma(S_t)^2 \mathrm{d}t \\
& = \left( \mu(S_t) \frac{\mathrm{d}\mu (S_t)}{\mathrm{d}x}  + \sigma(S_t)^2 \frac{1}{2} \frac{\mathrm{d}\mu(S_t)}{\mathrm{d}x^2}\right)\mathrm{d}t + \sigma(S_t)\frac{\mathrm{d}\mu (S_t)}{\mathrm{d}x}\mathrm{d}W_t
\end{split}
\end{equation}

since $\mu$ only depends on $S_t$ and not on $t$. 
In the same manner: 
\begin{align}
\mathrm{d}\sigma(S_t) &= \left( \mu(S_t) \frac{\mathrm{d}\sigma(S_t)}{\mathrm{d}x} + \sigma(S_t)^2 \frac{1}{2} \frac{\mathrm{d}\sigma(S_t)}{\mathrm{d}x^2}\right)\mathrm{d}t + \sigma(S_t)\frac{\mathrm{d}\sigma (S_t)}{\mathrm{d}x}\mathrm{d}W_t
\end{align}

In the integral form, this leads to: 

\begin{align}
\mu_s & = \mu_t+ \int^s_t \left ( \mu_u'\mu_u + \frac{1}{2} \mu_u'' \sigma^2_u \right )du + \int^s_t \mu_u' \sigma_u dW_u \\
\sigma_s & = \sigma_t+ \int^s_t \left ( \sigma_u'\mu_u + \frac{1}{2} \sigma_u'' \sigma^2_u \right )du + \int^s_t \sigma_u' \sigma_u dW_u
\end{align}

We can then rewrite the differential in integral form, with new $\sigma_t$ and $\mu_t$.

\begin{equation} 
\begin{split}
    \label{eq:eq_milstein_1}
    S_{t+dt} = & S_t + \int^{t+dt}_t \left ( \mu_t+ \int^s_t \left ( \mu_u'\mu_u + \frac{1}{2} \mu_u'' \sigma^2_u \right )du + \int^s_t \mu_u' \sigma_u dW_u \right )ds \\
    & + \int^{t+dt}_t \left (\sigma_t+ \int^s_t \left ( \sigma_u'\mu_u + \frac{1}{2} \sigma_u'' \sigma^2_u \right )du + \int^s_t \sigma_u' \sigma_u dW_u \right )dW_s
\end{split}
\end{equation}

For this approximation, we ignore terms with any higher order than $\mathcal{O}(dsdu) = \mathcal{O}(dt^2)$. Since $dW_udW_s = \mathcal{O}(dt)$ and $dudW_s = \mathcal{O}(dt^{\frac{3}{2}})$ Equation \ref{eq:eq_milstein_1} becomes: 

\begin{equation} 
\begin{split}
\label{eq:eq_milstein_2}
    S_{t+dt} = & S_t + \mu_t \int^{t+dt}_t ds + \sigma_t \int^{t+dt}_t dW_s + \int^{t+dt}_t \int^s_t \sigma_u' \sigma_u dW_u dW_s
\end{split}
\end{equation} 
We can again apply Euler discretisation the last term of Equation \ref{eq:eq_milstein_2}:
\begin{equation} 
\begin{split}
    \int^{t+dt}_t \int^s_t \sigma_u' \sigma_u dW_u dW_s & \approx \sigma_t' \sigma_t \int^{t+dt}_t \int^s_t dW_u dW_s \\
    & = \sigma_t' \sigma_t \int^{t+dt}_t \left ( W_s - W_t) \right )dW_s \\
    & = \sigma_t' \sigma_t \left ( - W_{t+dt}W_t + W_t^2 + \int^{t+dt}_t W_s dW_s \right)
\end{split}
\end{equation} 

Thanks to Ito's Lemma, we can derive: 

\begin{align}
\int^{t+dt}_t W_s dW_s &= \frac{1}{2}W_{t+dt}^2 - \frac{1}{2}W_t^2 + \frac{1}{2}dt
\end{align}

Recall that $W_{t+dt} - W_t$, has the same distribution as $\text{Z} \sim N(0,dt)$. Finally: 
\begin{equation} 
\begin{split}
\int^{t+dt}_t \int^s_t \sigma_u' \sigma_u dW_u dW_s & \approx \frac{1}{2}\left( (W_{t+dt} - W_t)^2 - dt \right ) \\
& = \frac{1}{2}dt \left (Z^2 - 1 \right )
\end{split}
\end{equation} 

We substitute this expression in equation \ref{eq:eq_milstein_2} to obtain the Milstein discretisation of the SDE :

\begin{align}
\label{eq:final_mil}
    S_{t+dt} & = S_t + \mu_tdt + \sigma_t\sqrt{dt}Z+ \frac{1}{2}\sigma' \sigma dt (Z^2 - 1)
\end{align}
The Milstein method adds a corrective term to Euler discretisation $\frac{1}{2}\sigma_t' \sigma_t dt (Z^2 - 1)$. 
We can now apply the Milstein discretisation to Heston model. Recall that Heston model is described by the following equations: 
\begin{align}
\mathrm{d}S_t &= r S_t \mathrm{d}t + \sqrt{v_t} S_t \mathrm{d}W^{(1)}_{t}\\
\mathrm{d}V_t &= \kappa(\theta-V_t) \mathrm{d}t + \sigma \sqrt{V_t} \mathrm{d}W^{(2)}_{t}
\end{align}

If we discretise $\ln(S_t)$ instead of $S_t$, we obtain, by Ito's lemma:

\begin{align}
d\ln(S_t) &= (r - \frac{1}{2} V_t)dt + \sqrt{V_t} dW^{(1)}_{t}\\
\end{align}
We can set $\mu(S_t) = (r - \frac{1}{2} V_t)$ and $\sigma(S_t) = \sqrt{V_t}$ so $\sigma(S_t)' = 0$. According to equation \ref{eq:final_mil}: 
\begin{align}
\ln(S_{t+dt}) &= \ln(S_t) +  (r - \frac{1}{2} V_t)dt + \sqrt{V_tdt}Z_1 \\
S_{t+dt} &= S_t\exp \left ((r - \frac{1}{2} V_t)dt + \sqrt{V_tdt}Z_1 \right )
\end{align}
Thus, Milstein approximation adds no correction term to $S_t$ discretisation. Regarding the variance process $V_t$, we can set: $\mu(V_t) = \kappa(\theta - V_t)$ and $\sigma(V_t) = \sigma \sqrt{V_t}$ so that $\sigma(V_t)'=\sigma * \frac{1}{2\sqrt{V_t}}$.

Finally, variance process can be approximated with: 
\begin{align}
V_{t+dt} = V_t + \kappa (\theta - V_t)dt + \sigma\sqrt{V_t dt}Z_2 + \frac{1}{4}\sigma^2dt(Z_2^2 - 1)
\end{align}

A bias introduced by Milstein algorithm is the possibility to have negative values for $V_t$. Thus, a \textit{truncation} algorithm is used. At each iteration, we take $max(V_{t+dt}, 0)$ for $V_{t+dt}$ value. Now we have the complete Milstein approximation algorithm \ref{alg:Milstein} for estimating an asset price $S_T$ under Heston model with initial values $S_0, V_0$, in $N$ discretisation steps.

\begin{algorithm}
\caption{Milstein approximation under Heston model}
\label{alg:Milstein}
\begin{algorithmic}[1]
\State $\Delta t \gets \frac{T}{N}$
\State $t \gets 0$
\State $V \gets V_0$
\State $S \gets S_0$
\While {$t<T$}
\State Generate $Z_1, Z_2 \sim N(0,1)$;
\State $Z_2 \gets \rho Z_1 + \sqrt{1 - \rho^2} Z_2$
\State $S \gets S_t\exp \left ((r - \frac{1}{2} V)\Delta t + \sqrt{V_t \Delta t}Z_1 \right )$
\State $V \gets V + \kappa (\theta - V)\Delta t + \sigma\sqrt{V \Delta t}Z_2 + \frac{1}{4}\sigma^2\Delta t(Z_2^2 - 1)$
\State $V \gets \max(V,0)$
\State $t \gets t+\Delta t$
\EndWhile
\end{algorithmic}
\end{algorithm}

\subsection{Monte Carlo methods}
Monte Carlo methods are a class of computational algorithms that rely on repeated random sampling to obtain numerical results. These methods are widely used in a variety of fields, including physics, finance, engineering, and statistics. In essence, Monte Carlo methods involve simulating a large number of possible outcomes for a given system, using random numbers to generate the necessary inputs. By simulating many possible outcomes and averaging the results, Monte Carlo methods can provide accurate estimates of complex systems. The law of large numbers ensures the convergence of Monte Carlo estimation to the actual value which is estimated. The Monte Carlo estimator $\phi$ of a quantity $f$, which depends on many variables, is: 

\begin{equation}
    \varphi^{MC}(f) = \frac{1}{N} \sum^N_{i=1} f(x_i)
\end{equation}
where $(x_1, ..., x_i, ..., x_N)$ is a random sequence of possibles values for $f$ inputs. It has been proved that Monte Carlo estimators converges at a $\mathcal{O}(\frac{1}{\sqrt{N}})$ rate.

Monte Carlo methods are commonly used for option pricing, particularly in cases where the option has a complex payoff or where analytic solutions are not available. The method involves simulating a large number of paths of the underlying asset price using stochastic models, such as Heston model, and then computing the option payoff at the end of each path. These payoffs are then discounted back to the present using the risk-free interest rate, and the resulting values are averaged to obtain an estimate of the option price. The accuracy of the Monte Carlo method depends on the number of simulated paths used, as well as the accuracy of the stochastic process used to model the underlying asset price. Due to the number of paths needed to obtain a good estimation, Monte Carlo methods can be computationally intensive, but their flexibility and the easy implementation make them very popular in finance.

Assume that we want to use Monte Carlo method to price an European option under Heston model. A way to generate independent asset paths, such as Exact Simulation, or Milstein approximation has to be chosen. $N$ asset paths will be simulated under the chosen model, and the option payoff will be averaged from them. Such a Monte Carlo algorithm is detailed in algorithm \ref{alg:Euro_MC}.

\begin{algorithm}
\caption{Monte Carlo option pricing estimation}
\label{alg:Euro_MC}
\begin{algorithmic}[1]
\State{avgPayoff $\gets 0$}
\For{$k \gets 1$ to N\_path}
\State{$S_T \gets $Milstein($N\_steps$)} \Comment{See alg. \ref{alg:Milstein}}
\State{avgPayoff $\gets$ avgPayoff + $\max(S_T-K, 0)$} \Comment{European Payoff}
\EndFor
\State{avgPayoff $\gets$ avgPayoff/N\_Paths}
\State{Payoff $\gets e^{-rT}$ avgPayoff} \Comment{Discounted payoff}
\end{algorithmic}
\end{algorithm}
\begin{algorithm}
\end{algorithm}

\subsection{Pseudo random number generation}
Every Monte Carlo method relies on a robust randomness of samplings. It is crucial to cover a large sample of numbers, behaving as random and independent draws, in order to obtain a good Monte Carlo estimation. Drawing random numbers on a computer, where nothing is really stochastic, is a challenge addressed by Pseudo Random Number Generators (PRNG). A PRNG goal is to generate numbers that have the same properties that genuinely random sequences, while being deterministic. One of the most popular PRNG is Mersenne Twister algorithm, developed by Makoto Matsumoto and Takuji Nishimura in 1997.The most popular version of this algorithm, MT19937, offers several advantages:  

\begin{itemize}
    \item It has a period of $2^{19937} - 1$
    \item It is uniformly distributed on a high number of dimensions 
    \item It is faster that most of the similar randomness quality algorithms
\end{itemize}

MT19937 is the default PRNG in languages such as Python, Ruby, PHP, and is available in standard version of C++. 
\subsection{Quasi Monte Carlo methods}
\begin{figure}
    \centering
    \includegraphics[width=0.75\textwidth]{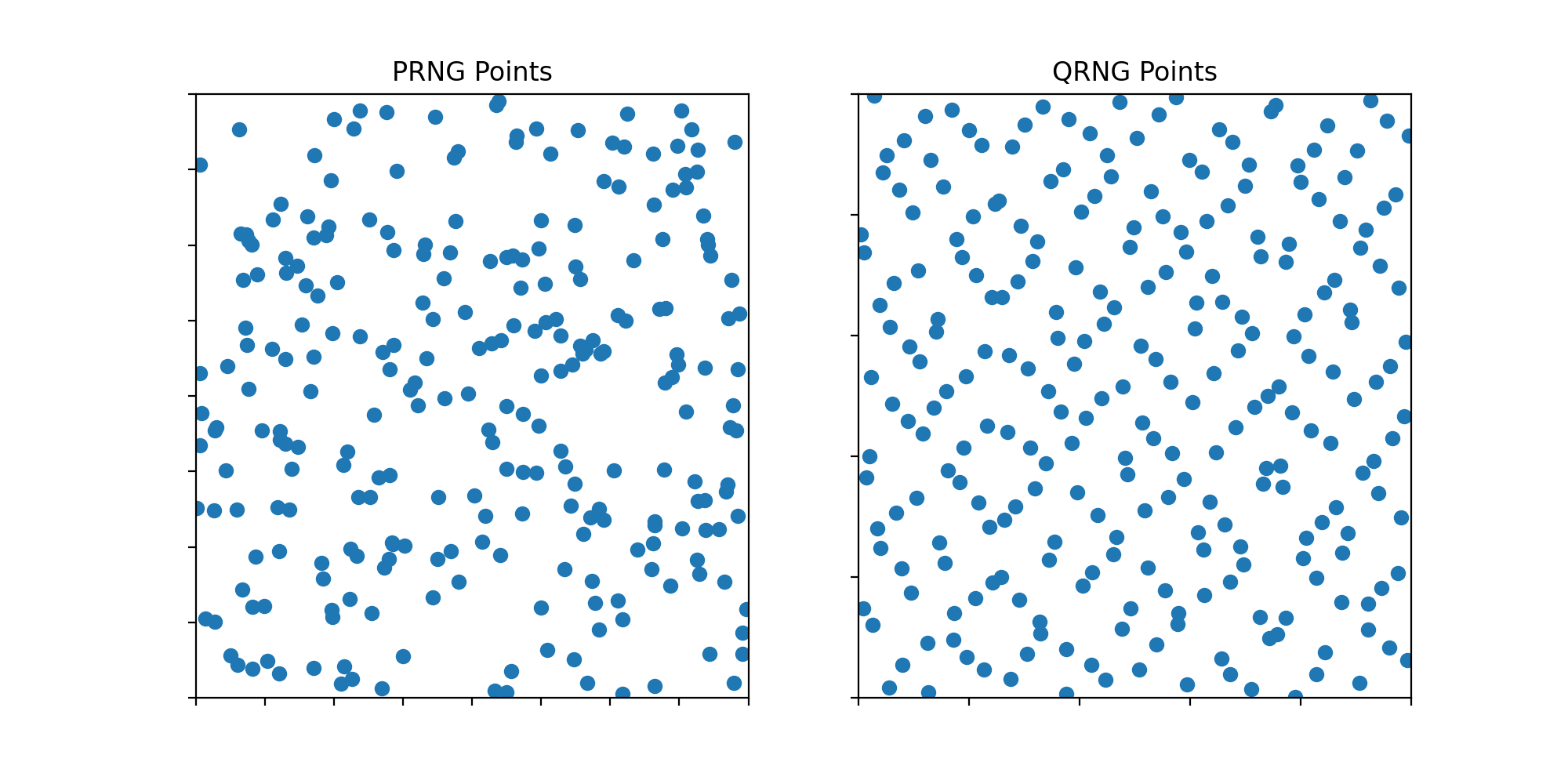}
    \caption{PRNG vs QRNG}
    \label{fig:mc_qmc}
\end{figure}

While PRNGs try to simulate a true random distribution, another way to sample random numbers for Montecarlo methods is often used: Quasi Random Number Generators (QRNG). Unlike PRNG, QRNG do not try to mimic a truly random number generation, but use evenly distributed numbers, called Low Discrepancy Sequences. Quasi Monte Carlo (QMC) methods use LDS, which goal is to sample point in a more uniform way. To illustrate this example, a sample of 256 points have been draught through PRNG and QRNG in figure \ref{fig:mc_qmc}. QMC offers a clear advantage in terms of speed of convergence. Whereas MC methods converge at a $\mathcal{O}(\frac{1}{\sqrt{n}})$, QMC methods converge at a speed of $\mathcal{O}(\frac{1}{n})$. The trade off is that for high dimension problems, a lot of samples is needed, because the error rate is $\mathcal{O}(\frac{log(n)^{d}}{n})$. Thus, if we consider a problem with 360 dimensions, the error would be $(\frac{log(10^9)^{360}}{10^9}) > 10^{360}$. QMC methods are unusuable in pratcice for high dimensions problems. 

Sobol sequences are a type of quasi-random sequence that are commonly used in quasi-Monte Carlo methods for numerical integration and option pricing. They were first introduced by the Russian mathematician Ilya Sobol in 1967 \cite{SOBOL196786}.

A Sobol sequence $\boldsymbol{x}_n$ in $d$ dimensions is a sequence of points in the unit hypercube $[0,1]^d$ that is constructed to have low discrepancy. The $n$th point in the sequence is obtained by applying a set of $d$ direction numbers $\boldsymbol{v}_1, \dots, \boldsymbol{v}_d$ to the integer $n$ in base $2$, with each digit representing a bit of $n$. Without covering too much details about the how Sobol sequences are computed, it worth notice that they satisfy two properties, namely $A$ and $A'$. 

\begin{itemize}
    \item A low-discrepancy sequence is said to satisfy Property A if for any binary segment (not an arbitrary subset) of the d-dimensional sequence of length 2d there is exactly one draw in each 2d hypercubes that result from subdividing the unit hypercube along each of its length extensions into half.
    \item A low-discrepancy sequence is said to satisfy Property A’ if for any binary segment (not an arbitrary subset) of the d-dimensional sequence of length 4d there is exactly one draw in each 4d hypercubes that result from subdividing the unit hypercube along each of its length extensions into four equal parts.
\end{itemize}

\subsection{Parallelisation on CUDA}
It is easily seen that Monte Carlo estimation, such as the one described in Algorithm \ref{alg:Milstein}, can be parallelised, because of the generation of independent paths. CUDA (Compute Unified Device Architecture) is a parallel computing platform and programming model developed by NVIDIA. It was first introduced in 2006 as a way to enable general-purpose computing on graphics processing units (GPUs), which were originally designed to handle computer graphics and video decoding. CUDA allows developers to write programs that can take advantage of the massive parallel processing power of modern GPUs to accelerate a wide range of computationally intensive tasks. Since its introduction, CUDA has become one of the most widely used platforms for GPU computing, with support for a wide range of programming languages and tools. CUDA programs are most of the time \textit{Heterogeneous}, which means a part of the program is executed on CPU (the \textit{host}) while another part is executed on GPU (the \textit{device}).

\begin{figure}
    \centering
    \includegraphics[width=0.65\textwidth]{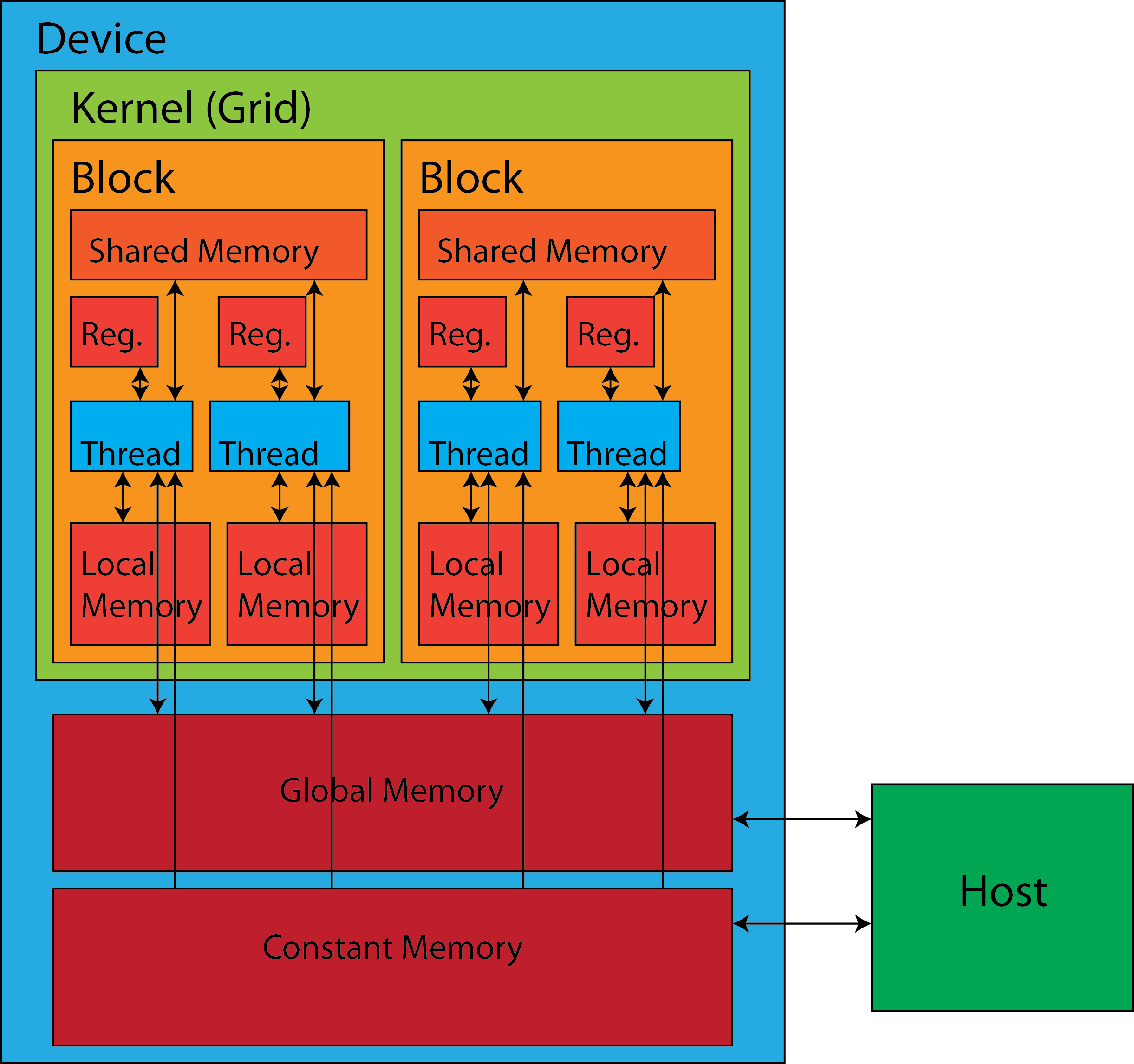}
    \caption{CUDA architecture obtained from \url{https://cvw.cac.cornell.edu/gpu/memory_arch}}
    \label{fig:cuda_arch}
\end{figure}

When using CUDA to parallelize a task, the work is divided into a grid of blocks, and each block is further divided into threads. The number of threads per block and the number of blocks per grid can be specified by the programmer. Each thread runs the same code, but operates on a different piece of data. The threads within a block can communicate with each other and share data using shared memory, while threads in different blocks cannot communicate directly. The CUDA software manages the allocation of work to threads and blocks and the synchronization of their operations. The architecture of CUDA workflow is illustrated in figure \ref{fig:cuda_arch}.

One of the main limitations of CUDA is its memory capacity. In CUDA programming, the memory is divided into two parts: global memory and shared memory. The global memory is a large, slow memory that is accessible to all threads in a grid, while the shared memory is a smaller, faster memory that is shared between threads in a block. However, the amount of shared memory is limited, and excessive use of global memory can significantly slow down the program. In addition, the transfer of data between the host and device memory can also be a bottleneck, especially when dealing with large amounts of data. 

\section{Experiment}

Our code can be found at \url{https://github.com/pierreia/QMCHestonGPU}. This code allows its user to price an European or Asian option under Heston model on both CPU and GPU, and to compute their Greeks in a flexible manner. 

In this experiment, the aim is to improve the efficiency of option Greeks estimation by implementing it on a GPU with an original approach. To establish the effectiveness of the GPU implementation, CPU baselines are first established. Subsequently, the calculation is implemented on the GPU and compared with the CPU implementation. The main objective of this experiment is to demonstrate that using a GPU for option Greeks estimation can result in significant computational advantages over traditional CPU-based methods. The experimental setup uses an Intel I5 CPU and an Nvidia GTX 1060Ti GPU. The parameters used for both European Options and Asian Options are presented in Table \ref{tab:heston}. All the numerical results obtained in a simulation are evaluated over 30 runs and reported as \textit{runs mean (std error)}.

\begin{center}
\begin{table}
\centering
\begin{tabular}{ |c|c|c|c|c|c|c|c|c| }
 \hline
 Variable & $\kappa$ & $\theta$ & $\sigma$ & $v_0$ & $T$ & $r$ & $S_0$ & $K$ \\
 \hline
 Value & 6.21 & 0.019 & 0.61 & 0.010201 & 1 & 0.0319 & 100 & 100 \\
 \hline
\end{tabular}
\caption{Heston model parameters}
\label{tab:heston}
\end{table}
\end{center}

\subsection{Baseline Implementation}
\begin{figure}
    \centering
    \includegraphics[width=0.45\textwidth]{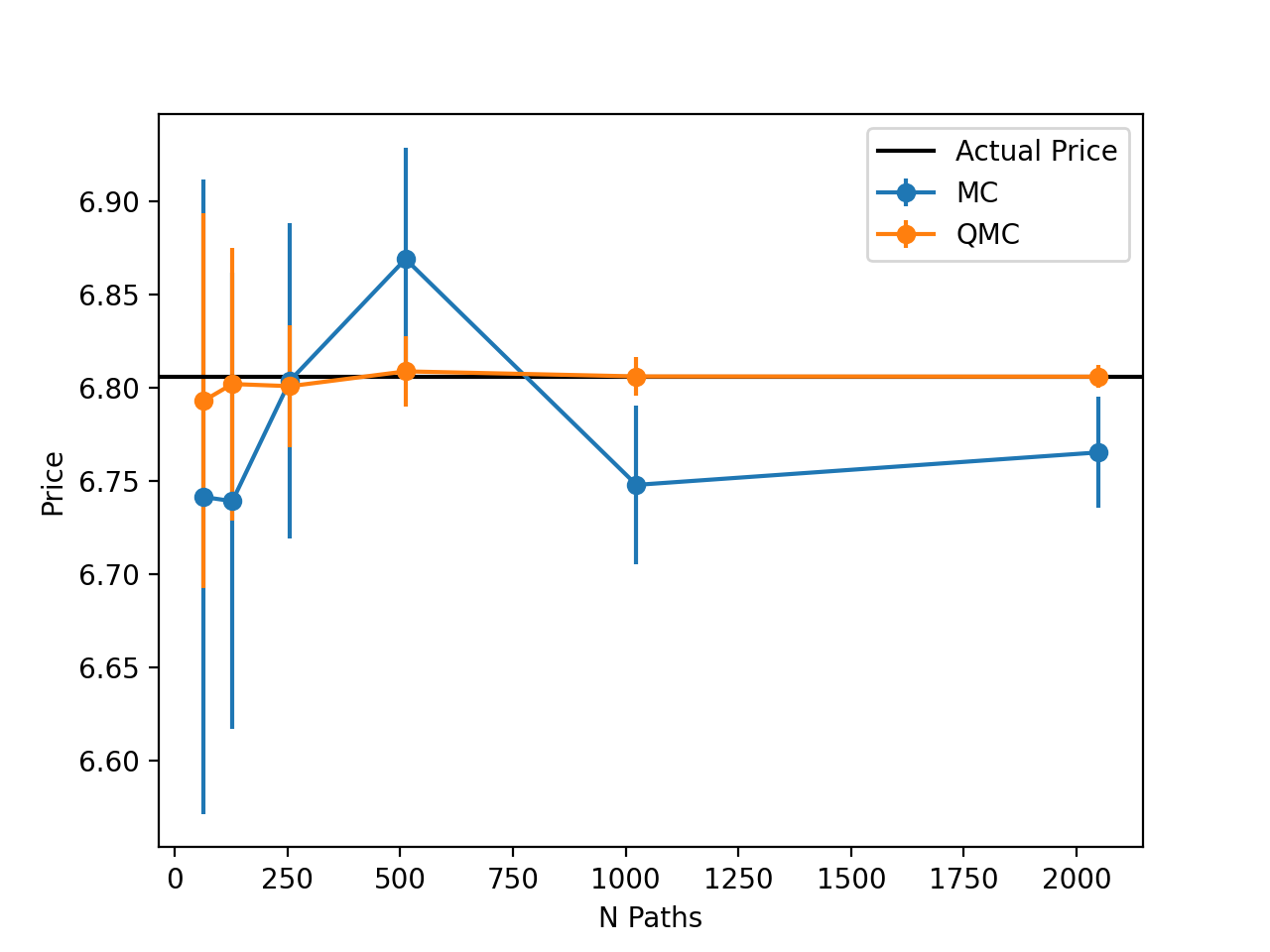}
    \includegraphics[width=0.45\textwidth]{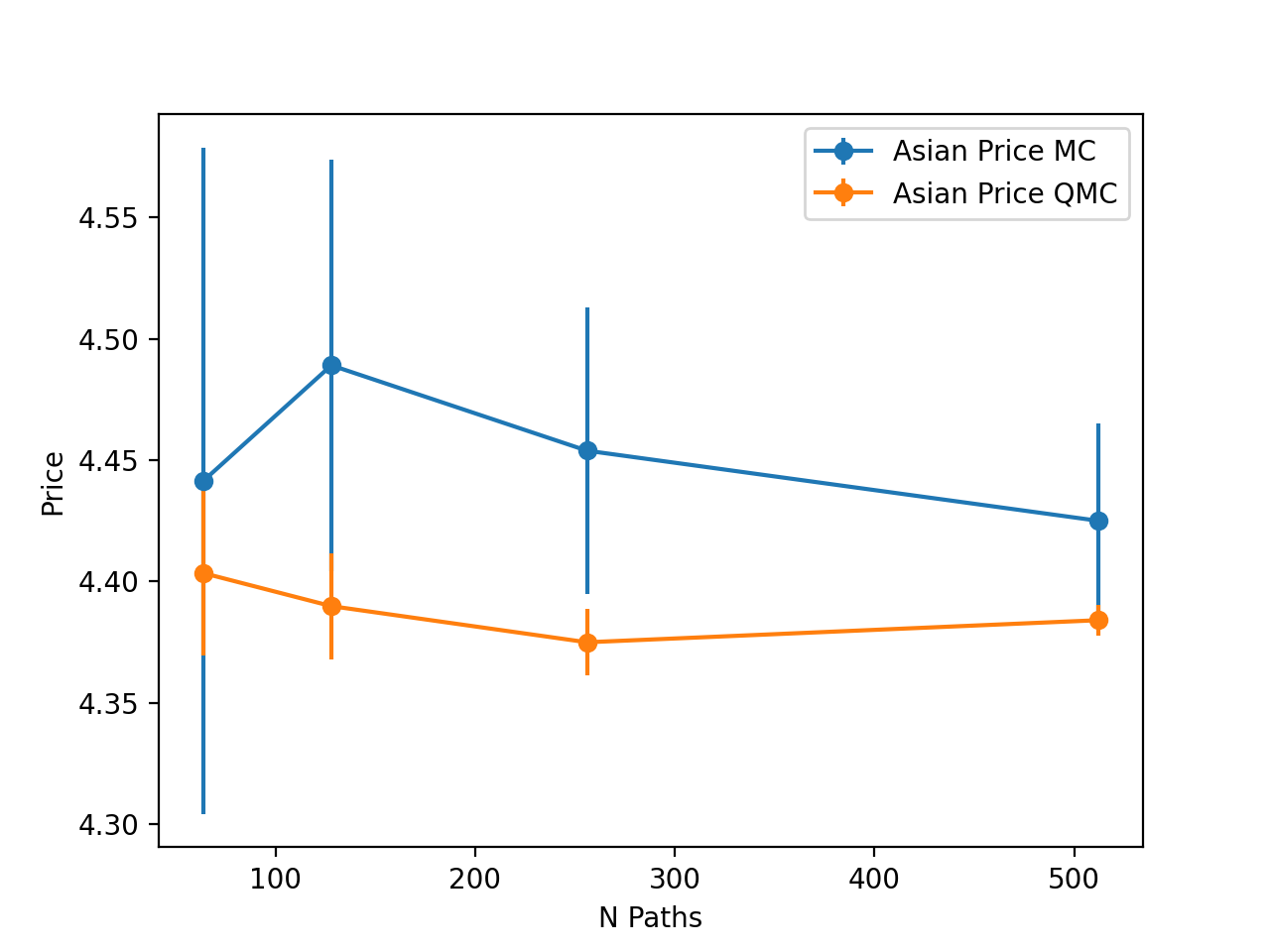}
    \caption{Exact Simulation results for European and Asian Options}
    \label{fig:exact_plot}
\end{figure}

In this part of the report, we present the baseline implementation used to establish the CPU performance of the option Greeks estimation. We used the exact Heston simulation implementation in C++ by Jan Baldeaux and Dale Roberts \cite{baldeaux2012quasimonte}, which was based on the work of Broadie and Kaya \cite{ExactBroadie}. This implementation serves as the benchmark for precision and speed of execution. As per the Monte Carlo method, multiple asset prices were generated using algorithm \ref{alg:exact}, and the price paths were averaged. The results were obtained by running 30 simulations for each parameter set of European and Asian options, as presented in Table \ref{tab:exact} and \ref{tab:exact_asian}, respectively, and Figure \ref{fig:exact_plot}. Since the exact simulation method used only three random samples per path, we could use the Quasi Monte Carlo method due to the small dimension of the problem. It significantly improves the simulation result.

\begin{center}
\begin{table}
\centering
\label{tab:exact}
    \begin{tabular}{|c|c|c|c|c|}
    
    \hline
        N Paths & Price MC & Price QMC & Time MC (ms) & Time QMC (ms)  \\
    \hline
64 & 6.7415 (0.1699) & 6.7931 (0.1004) & 225ms & 232ms \\
128 & 6.7393 (0.1222) & 6.8021 (0.0732) & 453ms & 459ms \\
256 & 6.8040 (0.0845) & 6.8009 (0.0325) & 909ms & 913ms \\
512 & 6.8691 (0.0597) & 6.8089 (0.0190) & 1820ms & 1821ms \\
1024 & 6.7480 (0.0423) & 6.8062 (0.0102) & 3630ms & 3657ms \\
2048 & 6.7654 (0.0299) & 6.8060 (0.0060) & 7277ms & 7284ms \\
        \hline
    \end{tabular}
\caption{Exact Simulation of European Option, real price = 6.8061}
\end{table}
\end{center}

\begin{center}
\begin{table}
\centering
\label{tab:exact_asian}
    \begin{tabular}{|c|c|c|c|c|}
    
    \hline
        N Paths & Price MC & Price QMC & Time MC (ms) & Time QMC (ms)  \\
    \hline
      64 &           4.4415 (0.1374)  &          4.4034 (0.0340) & 2108ms & 2117ms  \\
     128 &           4.4891 (0.0847)  &          4.3898 (0.0218) & 4208ms & 4226ms  \\
     256 &           4.4539 (0.0589) &          4.3749 (0.0138) & 8443ms  & 8442ms  \\
     512 &   4.4250 (0.0403) &   4.3840 (0.0064) & 16857ms  & 16892ms \\

         \hline
    \end{tabular}
\caption{Exact Simulation of Asian Option, Times = [0.25, 0.5, 0.75, 1]}
\end{table}
\end{center}

The process of parallelising the Exact Simulation algorithm (Algorithm \ref{alg:exact}) for the Heston model on a GPU was challenging due to its complexity. Initial attempts showed no speed gain because of reaching the GPU limit. Therefore, we turned to discretisation methods that are more amenable to parallelisation. However, discretisation methods require more random variables to generate a single asset path compared to the exact method, and generating a large number of paths is necessary to ensure convergence to the actual asset price. Despite attempts to implement Quasi Monte Carlo methods for Milstein and Euler discretisation, the results were unsatisfactory in the high-dimensional setting. Instead, we propose a novel method to compute option Greeks, called Milstein GPU (MGPU), using the Milstein discretisation on a GPU. We implemented Algorithm \ref{alg:Euro_MC} using CUDA kernels and cuRAND for random number generation. Notably, we generated random variables within each kernel, so the total execution time of a path under the MGPU implementation includes random variable generation. 

\subsection{Results}

\subsubsection{Discretisation steps}
Discretisation steps are a crucial choice when dealing with Euler like approximations. Thus, impact of timesteps on option price estimation have been evaluated and reported in Table \ref{tab:Milstein_timesteps}. It appears that a number of timesteps above 32 does not improve the final result. 

\begin{table}
\label{tab:Euler_timesteps}
\center
\begin{tabular}{|c|c|c|c|c|}
\hline
N Timesteps & CPU Price & GPU Price & CPU Time (ms) & GPU Time (ms) \\
\hline
32 & 4.3813 (0.0344) & 4.3661 (0.0470) & 254 & 26 \\
64 & 4.3757 (0.0589) & 4.3695 (0.0355) & 513 & 27 \\
128 & 4.3906 (0.0482) & 4.3888 (0.0492) & 1022 & 24 \\
256 & 4.3780 (0.0365) & 4.3757 (0.0506) & 2043 & 34 \\
\hline
\end{tabular}
\caption{Milstein Asian Price Estimation, 32 000 Paths, Exact Sim Price = 4.3840}  
\end{table}

\begin{table}
\label{tab:Milstein_timesteps}
\center
\begin{tabular}{|c|c|c|c|c|}
\hline
N Timesteps & CPU Price & GPU Price & CPU Time (ms) & GPU Time (ms) \\
\hline
32 & 6.8122 (0.0417) & 6.7875 (0.0379) & 271 & 32 \\
64 & 6.7939 (0.0415) & 6.8173 (0.0463) & 540 & 26 \\
128 & 6.8012 (0.0341) & 6.7874 (0.0419) & 1079 & 29 \\
256 & 6.7950 (0.0431) & 6.7882 (0.0316) & 2156 & 37 \\
\hline
\end{tabular}
\caption{Milstein European Price, 32 000 Paths, Real Price = 6.8061}  
\end{table}

\subsubsection{Number of Paths}
When using Monte Carlo methods, the estimated result will tends to the actual value as the number of simulations increase. However, trade off have to be found in order to get decent computation time and usable programs. To find a trade off, we compared different number of paths generated. Results were reported in Tables \ref{tab:paths_euro} and \ref{tab:paths_asian}. More than the estimated price, the number to be observed is the standard error. It appears that it does not increase after 16000 paths generated.  

\begin{table}
\center
\label{tab:paths_euro}
\begin{tabular}{|c|c|c|c|c|}
\hline
N Paths & CPU Price & GPU Price & CPU Time (ms) & GPU Time (ms) \\
\hline
4000 & 6.7770 (0.1359) & 6.7734 (0.0996) & 127 & 6 \\
8000 & 6.8032 (0.1024) & 6.7889 (0.0667) & 254 & 11 \\
16000 & 6.8165 (0.0579) & 6.8033 (0.0461) & 508 & 18 \\
32000 & 6.8115 (0.0462) & 6.8185 (0.0484) & 1016 & 28 \\
\hline
\end{tabular}
\caption{Milstein European Price Estimation, 128 Timesteps, Real Price = 6.8061}  
\end{table}

\subsubsection{Asian Options}

\begin{table}
\label{tab:paths_asian}
\center
\begin{tabular}{|c|c|c|c|c|}
\hline
N Paths & CPU Price & GPU Price & CPU Time (ms) & GPU Time (ms) \\
\hline
4000 & 4.3874 (0.0583) & 4.3789 (0.0711) & 134 & 6 \\
8000 & 4.3984 (0.0636) & 4.3902 (0.0411) & 268 & 12 \\
16000 & 4.3907 (0.0266) & 4.3853 (0.0453) & 536 & 19 \\
32000 & 4.3779 (0.0284) & 4.3846 (0.0304) & 1074 & 28 \\
\hline
\end{tabular}
\caption{Milstein Arithmetic Asian Price Estimation, 128 Timesteps, Times = [0.25, 0.5, 0.75, 1.]}  
\end{table}

\subsubsection{Greeks}

In their work \cite{inproceedings}, Broadie and Kaya computed Greeks for both European and Asian options under an Exact Simulation with Heston model. These results are referred as \textit{Ex Delta} and \textit{Ex Rho}. More over, they provided actual values for European Option Greeks, referred as \textit{Delta} and \textit{Rho}. We used MGPU and CPW as presented in Equations \ref{eq:euro_pw} and \ref{eq:asian_pw} to compute Greeks showed in Table \ref{tab:greeks}.
The speed up gained from MGPU method on Exact Simulation is especially high for Asian Options. This is due to the fact that Milstein discretisation will discretise asset path to get its price, thus, computing the mean does not add any computational cost to the price generation process. On the contrary, the Exact Simulation does not discretise an asset path by default, so this operation needed to compute Asian option price (and so Greeks) multiplies the execution time compared to an European option. Nevertheless, surprising results were obtained on Rho estimation. 

\begin{figure}
\label{fig:greeks}
    \centering
    \includegraphics[width=1.\textwidth]{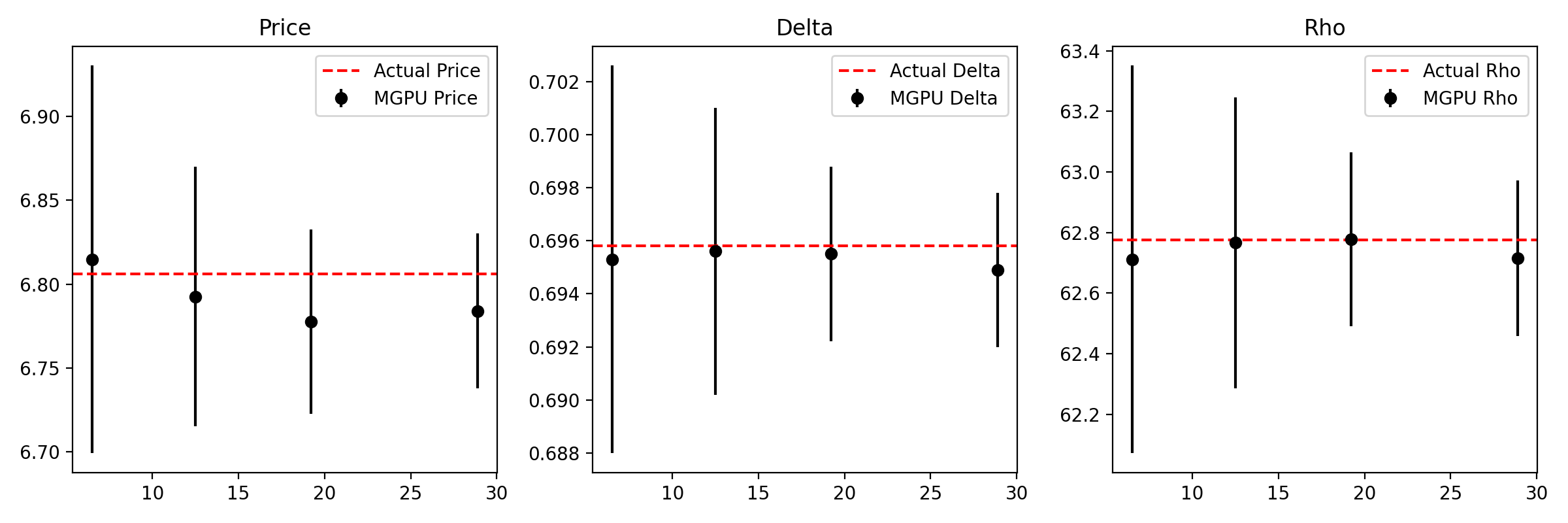}
    \includegraphics[width=1.\textwidth]{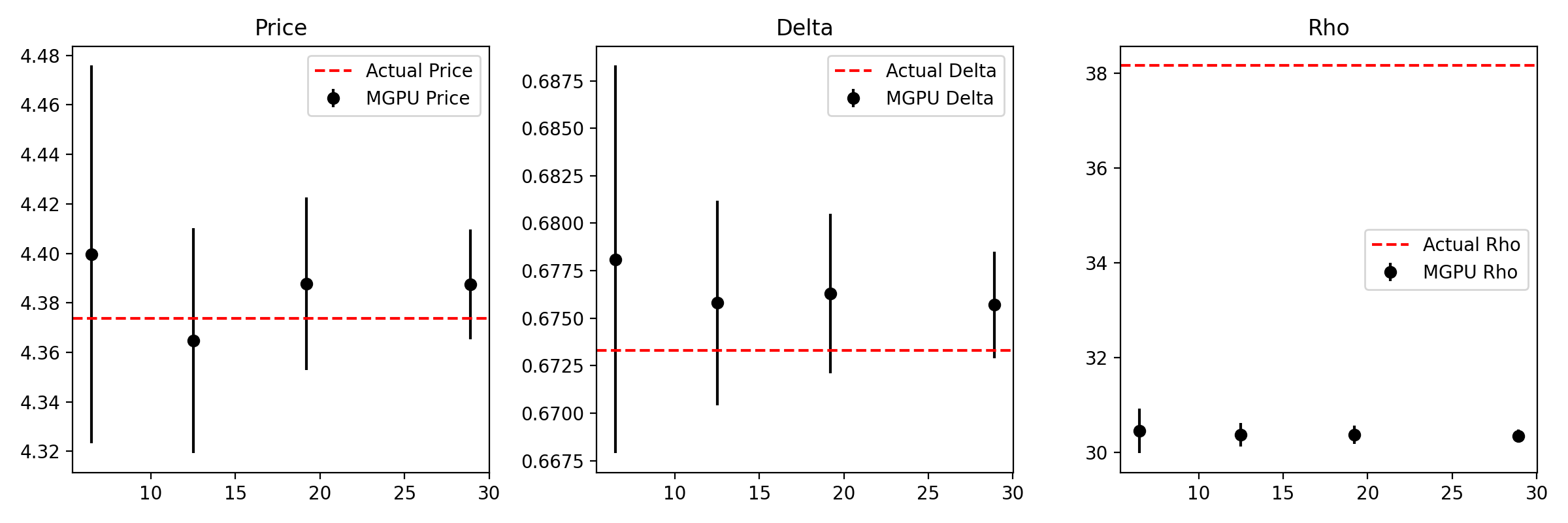}
    \caption{MGPU vs Time (ms) results}
    \label{fig:global_results}
\end{figure}

However, the results for Rho estimation were not as expected. While the Milstein method on CPU gives results (36.7856) similar to Exact Simulated Rho (38.1664), the GPU algorithm provides disappointing results, as presented in Figure \ref{fig:greeks}.

\begin{table}
\label{tab:greeks}
\center
\begin{tabular}{|c|c|c|c|c|c|c|c|}
\hline
 Option & Delta &  Ex Delta & GPU Delta & Rho & Ex Rho &  GPU Rho & Speed Up \\
\hline
European & 0.6958 & 0.6952 & 0.6956 & 62.7752 & 62.7148 & 62.7609 & 36x \\
Asian & N/A & 0.6733  & 0.6753 & N/A & 38.1664 & 30.3149 & 176x  \\
\hline
\end{tabular}
\caption{European and Asian Options Greeks}  
\end{table}

\subsection{Conclusion}
In this study, we have presented a novel implementation of the Heston model simulation on GPU using the Milstein discretisation. Our implementation, MGPU, achieved significant speedup over the Exact Simulation CPU implementation, especially for the computation of Asian option prices. Additionally, we presented the challenges associated with implementing the exact simulation method on GPU and proposed the Milstein discretisation method as a viable alternative.

Furthermore, we computed Greeks using MGPU, and found that MGPU performed well for Delta estimation. However, we observed a discrepancy in Rho estimation using MGPU, which requires further investigation.

In conclusion, our study demonstrates the potential benefits of GPU acceleration for Heston model simulation and highlights the the potential of discretisation methods for financial problems thanks to modern computational power. 

\subsubsection{Future Work}
While Exact Simulation method is not ideal to parallelise, some alternatives, which are based on Exact Simulation, may be more adapted to a GPU implementation. These methods, such as Andersen, significantly cut the computational cost while keeping great performance. It would be interesting to implement such a method on GPU in order to compare the results with our proposed solution.

\bibliographystyle{plain}

\end{document}